\documentclass[12pt,sort&compress]{iopart}
\pdfoutput=1 
\usepackage{graphicx}
\usepackage{color,amssymb}
\usepackage{cite}

\newcommand{\erfc}[1]{\mathrm{erfc}\! \left( {#1} \right)}

\begin{document}
\title{Dynamical properties of single-file diffusion}

\author{P.~L.~Krapivsky}
\address{Department of Physics, Boston University, Boston, MA 02215, USA}
\ead{paulk@bu.edu}

\author{Kirone Mallick}
\address{Institut de Physique Th\'{e}orique CEA, IPhT, F-91191 Gif-sur-Yvette, France}
\ead{kirone.mallick@cea.fr}

\author{Tridib Sadhu \footnote{Present address:
 Philippe Meyer Institute for Theoretical Physics, Physics Department, Ecole Normale Sup\'{e}rieure, 24 rue Lhomond, 75231 Paris Cedex 05 - France}}
\address{Institut de Physique Th\'{e}orique CEA, IPhT, F-91191 Gif-sur-Yvette, France}
\ead{tridib.sadhu@lpt.ens.fr}

\begin{abstract}
We study the statistics of a tagged particle in single-file diffusion,
a one-dimensional interacting infinite-particle system in which the
order of particles never changes. We compute the two-time
correlation function for the displacement of the tagged particle for
an arbitrary single-file system. We also discuss single-file analogs
of the arcsine law and the law of the iterated logarithm
characterizing the behavior of Brownian motion. Using a macroscopic
fluctuation theory we devise a formalism giving the cumulant
generating functional. In principle, this functional contains the full
statistics of the tagged particle trajectory---the full single-time
statistics, all multiple-time correlation functions, etc.  are merely
special cases. 
\end{abstract}

\vspace{2pc}

\submitto{A special issue of J Stat. Mech}

\maketitle

\section{Introduction}

Single-file diffusion refers to an infinite system of interacting particles diffusing on a line, or hopping on a one-dimensional lattice, which cannot overtake each other. Such crowded motion arises in many quasi one-dimensional systems (see e.g. \cite{Richards1977,KARGER1992,KUKLA1996,Li2009,Das2010,Siems2012}). The name `single-file' is particularly apt if one studies the evolution of a tagged particle. (The tagged particle is indistinguishable from the host particles, it merely carries a mark.) The behavior of the tagged particle is drastically different from Brownian motion \cite{Harris1965,Arratia1983,Rodenbeck1998,Leibovich2013,KMS2014,KMS2015,Sethuraman2013,Hegde2014}. This has also been demonstrated is many experimental realizations of single-file diffusion \cite{Wei2000,Lutz2004,Lin2005}. The goal of this article is to highlight these differences by comparing dynamical properties of the tagged particle in single-file diffusion (known and those we compute below) with corresponding features of Brownian motion. 

We limit ourselves to the simplest version of single-file diffusion with uniform initial density. Denote by $X(t)$ the displacement of the tagged particle. There is no drift in the case when the density profile is uniform, so the average displacement vanishes, $\langle X\rangle=0$, if we set $X(0)=0$. Higher odd moments also vanish: $\langle X^3 \rangle=\langle X^5 \rangle=\ldots=0$. The second moment exhibits a sub-diffusive growth
\begin{equation}
\label{SD_1d}
\langle X^2 \rangle = 2\mathcal{D}(\rho)\,\sqrt{t} 
\end{equation}
for large $t$.
This remarkable behavior was originally established 50 years ago \cite{Harris1965} for the simplest single-file system composed of Brownian particles with hard-core repulsion. Intriguingly, the same sub-diffusive growth \eref{SD_1d} applies to all single-file systems, each one is parametrized by a single amplitude $\mathcal{D}(\rho)$ depending on the (uniform) particle density $\rho$. This amplitude can be expressed \cite{KMS2014} through the two fundamental transport coefficients, the diffusion coefficient $D(\rho)$ and mobility $\sigma(\rho)$, underlying the large-scale hydrodynamic behavior of single-file diffusion:
\begin{equation}
\label{var_X}
\mathcal{D}(\rho) = \frac{\sigma(\rho)}{\rho^2\sqrt{4\pi D(\rho)}}
\end{equation}

For a random process without drift, the second moment does not generally characterize its probability distribution. In the case of single-file diffusion, however, Arratia proved \cite{Arratia1983} that it does and he showed that the probability distribution of the tagged particle position converges to a Gaussian \cite{Arratia1983}. This striking result has a strong psychological influence, one starts to think that the behaviors of the tagged particle and Brownian motion are qualitatively similar and after replacing $t$ in formulas for Brownian motion by $\sqrt{t}$ one would recover correct results for single-file diffusion. This intuition is wrong as we shall demonstrate by comparing a few basic features of Brownian motion with their single-file analogs.

The crucial general distinction between Brownian motion and single-file diffusion is that for the latter one must carefully define the averaging. The dynamics is stochastic, so we obviously perform the averaging over different realizations of dynamics. Another source of randomness  is hidden in the initial conditions. Single-file diffusion is an infinite particle system, so even if the tagged particle is initially at the origin, $X(0)=0$ as we always assume, the initial positions of other particles affect the behavior of the tagged particle. If we consider a fixed initial condition which is macroscopically uniform (e.g., the equidistant distribution), instead of \eref{SD_1d} one finds \cite{Leibovich2013,KMS2014,KMS2015}
\begin{equation}
\label{SD_det}
\langle X^2 \rangle_{\rm det} = \mathcal{D}(\rho)\,\sqrt{2t}
\end{equation}
The usually cited sub-diffusive growth \eref{SD_1d} tacitly presumes that we additionally perform the averaging over all (macroscopically uniform) initial conditions. It is not surprising that the behaviors \eref{SD_1d}  and \eref{SD_det} are different for small and moderate times. In the long-time limit, however, one anticipates that an equilibrium distribution of particles is reached even if we start with a very non-equilibrium, e.g. equidistant, initial distribution. This seemingly implies that the initial condition will be `forgotten' and the mean-square displacement (and any other characteristics, e.g., $\langle X^4 \rangle$) should not depend on the averaging over the initial conditions. What looks even more striking is that for {\em any} initial condition the mean-square displacement has the same asymptotic behavior, Eq.~\eref{SD_det}, yet after averaging we obtain the result which is $\sqrt{2}$ times larger, Eq.~\eref{SD_1d}. 
\begin{figure}
\begin{center}
\includegraphics[width=0.5\textwidth]{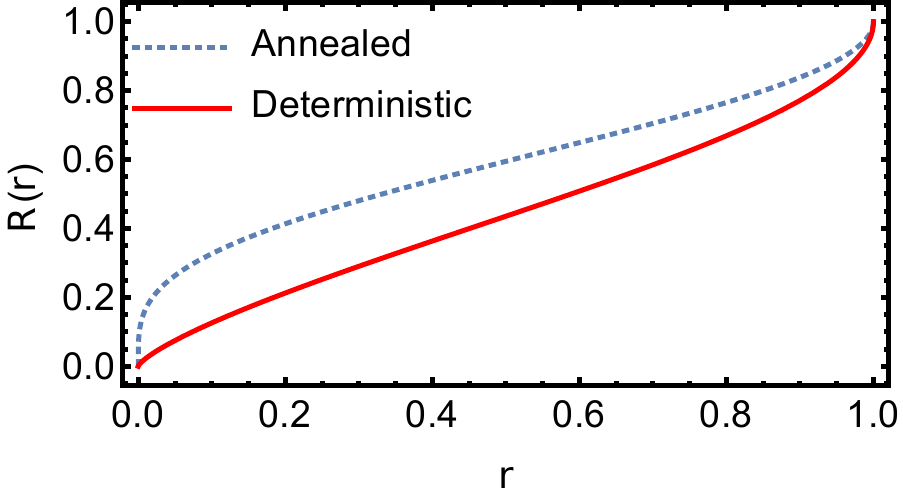}
\end{center}
\caption{Plot of $R(r)$ in \eref{eq:Rr0} for the two initial settings.  The function $R(r)$ does not depend on $\mathcal{D}(\rho)$ and increases monotonically from $0$ to $1$ with increasing $r$. \label{fig:plot1}}
\end{figure}

In this paper we study a few basic properties of single-file diffusion that have simple counterparts in the context of Brownian motion. One such property is the two-time correlation function. The neat formula
\begin{equation}
\label{BM:2T}
\langle x(t_1) x(t_2)\rangle = 2D\,\min(t_1, t_2)
\end{equation}
for the two-time correlation function encapsulates the major properties of Brownian motion, e.g., it can be derived from the equal-time correlation function $\langle x^2 \rangle = 2Dt$ using Markov property. For single-file diffusion, we shall obtain (for large $t_1$ and $t_2$)
\begin{equation}
\label{SF:2T}
\langle X(t_1) X(t_2)\rangle_{\rm ann} = \mathcal{D}(\rho)\left[ \sqrt{t_1}+\sqrt{t_2}-\sqrt{|t_2-t_1|} \right]
\end{equation}
if we perform the averaging over the initial conditions and
\begin{equation}
\label{SF:2T_det}
\langle X(t_1) X(t_2)\rangle_{\rm det} = \mathcal{D}(\rho)\left[ \sqrt{t_1+t_2}-\sqrt{|t_2-t_1|} \right]
\end{equation}
for a fixed initial condition. The derivation of Eqs.~\eref{SF:2T}--\eref{SF:2T_det} for an {\em arbitrary} single-file system is one of our major results. To illustrate the difference between the two settings we plot (\fref{fig:plot1}) the ratio 
\begin{equation}
R(r)=\frac{\langle X(t_1)X(t_2)\rangle}{\sqrt{\langle X^2(t_1)\rangle\langle X^2(t_2)\rangle}}\,, \qquad  r=\frac{t_1}{t_2}\,.
\label{eq:Rr0}
\end{equation}
This quantity is also known as the Pearson correlation coefficient.

Intriguingly \cite{KrugKallabis1997}, one can deduce \eref{SF:2T} from \eref{SF:2T_det}. Indeed, let us allow the evolution to proceed up to time $T$, and start measuring the displacement of the tagged particle after that moment. Denote by $T+t_1$ and $T+t_2$ two later times and by
\begin{equation}
\label{Yt}
Y(t_j)=X(T+t_j)-X(T), \quad j=1, 2
\end{equation}
the proper displacements. The two-time correlation function is
\begin{eqnarray*}
\langle Y(t_1)Y(t_2)\rangle&=&\langle X(T+t_1)X(T+t_2)\rangle_{\rm det}
+\langle X^2(T)\rangle_{\rm det}\\
&-&\langle X(T+t_1)X(T)\rangle_{\rm det} - \langle X(T+t_2)X(T)\rangle_{\rm det}
\end{eqnarray*}
Using \eref{SF:2T_det} we re-write $\langle Y(t_1)Y(t_2)\rangle$ as 
\begin{eqnarray*}
\frac{\langle Y(t_1)Y(t_2)\rangle}{\mathcal{D}(\rho)}&=&\sqrt{2T+t_1+t_2}-\sqrt{|t_1-t_2|} + \sqrt{2T}\\
&-&\sqrt{2T+t_1}+\sqrt{t_1}-\sqrt{2T+t_2}+\sqrt{t_2}
\end{eqnarray*}
We now take the 
\begin{equation}
\label{limits}
\frac{t_1}{T}\to 0, \quad \frac{t_2}{T}\to 0
\end{equation}
limit and find 
\begin{equation}
\label{YY}
\langle Y(t_1)Y(t_2)\rangle = \mathcal{D}(\rho)\left[ \sqrt{t_1}+\sqrt{t_2}-\sqrt{|t_2-t_1|} \right]
\end{equation}
Thus we do not need to average over initial conditions: Performing the shift of time \eref{Yt} and moving far ahead, as in   Eq.~\eref{limits}, we effectively put the system into equilibrium, so \eref{YY} should be the same as \eref{SF:2T} and indeed it is. Needless to say, \eref{SF:2T} and \eref{SF:2T_det} reduce to \eref{SD_1d} and \eref{SD_det} for equal times and this explains why the amplitudes in \eref{SD_1d}  and \eref{SD_det} are different, and why the difference is universal by a factor $\sqrt{2}$.

Many quantities arising in the context of Brownian motion are described by the arcsine law. This law gives, e.g. the distribution of time that the Brownian particle spends to the right of the origin and the distribution of time at which the Brownian particle reaches maximum. Our second goal is to investigate their single-file analogs. Consider the trajectory $\{X(t): 0\leq t\leq T\}$ of the tagged particle and denote by 
\begin{equation}
T_+ = \textrm{Measure}(\{t\in [0,T]: X(t) > 0\})
\end{equation}
the total time that the particle spends to the right of the origin. This time scales linearly with $T$, so one expects the cumulative distribution to be the function of the ratio $T_+/T$, that is,  ${\rm Prob}(T_+\leq \tau T) = \mathbb{P}_1(\tau)$ in the $T\to\infty$ limit. Similarly denote by $T_*$ the time at which the displacement of the tagged particle reaches its maximum. This time $T_*$ is a random quantity which scales linearly with $T$. Hence we anticipate that ${\rm Prob}(T_*\leq \tau T) = \mathbb{P}_2(\tau)$ in the $T\to\infty$ limit. 

For Brownian motion, the distributions $\mathbb{P}_1(\tau)$ and $\mathbb{P}_2(\tau)$ coincide and  are known as arcsine distributions \cite{Ito1965,Feller1968,Morters2010}:
\begin{equation}
\label{eq:arcsin}
\mathbb{P}_1(\tau) = \mathbb{P}_2(\tau) = \frac{2}{\pi}\,\arcsin\!\big(\sqrt{\tau}\big)
\end{equation}

Finding the distributions $\mathbb{P}_1(\tau)$ and $\mathbb{P}_2(\tau)$ for single-file diffusion is an interesting challenge. We have computed one non-trivial moment $\langle \tau^2\rangle = \int_0^1 d\tau\,\tau^2 \mathbb{P}_1(\tau)$. The result is {\em independent} of the implementation of the single-file constraint, but depends on whether we perform or not the averaging over initial conditions, and both answers are smaller than the corresponding result $\langle \tau^2\rangle = \frac{3}{8}$ for Brownian motion. 

The third characteristic  we study is the growth of the maximal displacement. Recall that for Brownian motion the typical spread grows as $\sqrt{Dt}$, but maximal fluctuations grow slightly faster. It is useful to think about the smallest asymptotic envelope that contains Brownian trajectories. Informally, one can write
\begin{equation}
\label{BM:log}
|x(t)|\leq  \sqrt{2\langle x^2(t)\rangle\,\ln(\ln t)}=\sqrt{4Dt\,\ln(\ln t)}
\end{equation}
This is often referred to as the \textit{law of the iterated logarithm} \cite{Ito1965,Morters2010}. The precise meaning of \eref{BM:log} is that $|x(t)|\leq C\sqrt{4Dt\,\ln(\ln t)}$ is asymptotically correct (i.e., valid for sufficiently large time) for any $C>1$, whereas for any $C<1$ any Brownian trajectory crosses the boundaries of envelope for arbitrarily large times. More formally \cite{Ito1965,Morters2010}, 
\begin{equation}
\label{BM:log+}
\lim_{t\to\infty}\sup \frac{x(t)}{\sqrt{2\langle x^2(t)\rangle\,\ln(\ln t)}} = 1, \quad
\lim_{t\to\infty}\inf \frac{x(t)}{\sqrt{2\langle x^2(t)\rangle\,\ln(\ln t)}} = -1.
\end{equation}

The extension to the single-file diffusion is rather straightforward. We shall argue that the analog of \eref{BM:log+} is
\begin{equation}
\label{SF:log+}
\lim_{t\to\infty}\sup \frac{X(t)}{\sqrt{\langle X^2(t)\rangle\,\ln t}} = 1, \quad
\lim_{t\to\infty}\inf \frac{X(t)}{\sqrt{\langle X^2(t)\rangle\,\ln t}} = -1.
\end{equation}
The choice of $\langle X^2(t)\rangle$ depends on whether we perform or not the averaging over initial conditions. 

Equal time and two-time correlation functions of the tagged particle displacement [Eqs.~\eref{SD_1d}, \eref{SD_det}, \eref{SF:2T}--\eref{SF:2T_det}] tell us a lot about single-file diffusion. Accordingly, one would like to compute multi-time correlations, and more generally the probability of an entire trajectory. For Brownian motion, the latter probability can be represented by a path integral. In this formalism, the probability of a trajectory is expressed as \cite{Kac1949,Satya2005} 
\begin{equation}
P[x]\sim \exp\!\left\{-A[x(\tau)]\right\}
\end{equation}
where the functional of the Brownian trajectory
\begin{equation}
A[x(\tau)]=\frac{1}{4D}\int_{0}^{t}d\tau \left( \frac{dx}{d\tau} \right)^2
\end{equation}
is known as action as it is formally identical to the action of a classical free particle. 

For single-file diffusion, there is no comparably simple formula for the action. A macroscopic fluctuation theory (MFT),  a recently developed systematic approach to non-equilibrium fluctuations in diffusive systems (see \cite{Bertini2007,Derrida2007,Bertini2014} for a review), allows one to proceed. Using MFT we have shown \cite{KMS2014} how to analyze the full statistics of the tagged particle position in a general single-file diffusion. Here we extend the formalism to assign probability weight to an entire trajectory of the tagged particle. In particular, we study a Legendre transform of the action and express it in terms of solution of a variational problem. At a formal level, the analysis boils down to solving Euler-Lagrange equations corresponding to the variational problem. A series expansion of the Legendre transform generates all the multi-time cumulants of tagged particle position.

The advantage of our approach is its generality, viz. it applies to all single-file systems. The microscopic details of the system, e.g. the inter-particle interaction and the size of the particles, appear at the macroscopic scale only in terms of two transport coefficients $D(\rho)$ and $\sigma(\rho)$. It is still extremely difficult to solve the Euler-Lagrange equations for an arbitrary single-file system, essentially the only fully solvable case corresponds to a single-file system composed of Brownian particles. Less complete results, however, can be established for an arbitrary single-file system. 
Specifically, we shall describe a perturbation scheme which allows, in principle, to compute the multi-time correlations order by order. Using this method we derive general formulas \eref{SF:2T}--\eref{SF:2T_det} for the two-time correlation function. 

One important result that emerges from our macroscopic analysis is that the multi-time cumulants of tagged particle position, $k_n(t_1,\cdots,t_n)=\langle X(t_1)\cdots X(t_n)\rangle_c$, scale sub-diffusively. More precisely, all multi-time cumulants scale as
\begin{equation}
k_n(t_1,\cdots,t_n)\sim \sqrt{T}
\label{eq:kn scaling}
\end{equation}
when $T={\rm max}(t_1,\ldots,t_n)$ is large. Correlation functions $\langle X(t_1)\cdots X(t_n)\rangle$ can be expressed in terms of the cumulants. Using these expressions one finds that for $n\geq 2$, the asymptotic behavior of the $n$-time correlation function is determined by the two-time cumulant (the one-time cumulant is zero due to symmetry), the contribution from other cumulants can be ignored. For example 
\begin{eqnarray}
\fl \quad  \langle X(t_1)X(t_2)X(t_3)X(t_4)\rangle= k_2(t_1,t_2)k_2(t_3,t_4)+k_2(t_1,t_3)k_2(t_2,t_4)+k_2(t_1,t_4)k_2(t_2,t_3)\cr
\qquad \qquad \qquad \qquad +k_4(t_1,t_2,t_3,t_4)
\end{eqnarray}
and using \eref{eq:kn scaling} we see that the $k_4$ term is negligible compared to the rest in the $T\to\infty$ limit.
 
The fact that the asymptotic behavior is expressed through the two-time cumulants implies an important feature of the joint distribution of tracer position at multiple times, viz. this joint distribution is asymptotically a multivariate Gaussian distribution:
\begin{equation}
P(X_1,\cdots,X_n)\simeq \frac{1}{\sqrt{(2\pi)^n\vert {\bf \Sigma}\vert}} e^{-\frac{1}{2}\bf X^t \Sigma^{-1} \bf X}
\label{eq:multi Gaussian}
\end{equation}
Here $X_j=X(t_j)$ and we used a vector notation ${\bf X}\equiv (X_1,\cdots,X_n)$ with $\bf X^t$ being the transpose of $\bf X$. Further, $\bf \Sigma$ is an $n\times n$ covariance matrix with $\Sigma_{ij}=\langle X(t_i)X(t_j)\rangle$ and $\vert {\bf \Sigma} \vert$ denotes determinant of $\bf \Sigma$. (The covariance matrix depends on the setting, one should use \eref{SF:2T} and \eref{SF:2T_det} for the annealed and the deterministic cases, respectively.) The asymptotically Gaussian distribution \cite{Arratia1983} of $X(t)$ is a special case of \eref{eq:multi Gaussian}. 

A direct consequence of \eref{eq:multi Gaussian} is that all multi-time correlations can be expressed in terms of the two-time correlation using the Wick's theorem (or the Isserlis' theorem).

The outline of this work is as follows. In \sref{sec:arcsine}, we consider  the time spent by the tagged particle to the right of the origin and compute the second and third moments. In \sref{sec:logarithm}, we give an heuristic derivation of the `law of the simple logarithm', Eq.~\eref{SF:log+}.  The MFT framework  is extended to the case of a functional of the entire trajectory in \sref{sec:mft} and the formalism is used to derive the two-time correlation functions  (\ref{SF:2T}, \ref{SF:2T_det}) in \sref{sec:2time}. We summarize our findings in \sref{sec:summary}.

\section{The arcsine law}
\label{sec:arcsine}

Consider a Brownian motion $x(t)$ which evolves during the time interval $0 \le t \le 1$. Let $\tau$ be the time spent by this $x(t)$ on the positive side i.e.,
\begin{equation}
\tau  = \int_0^1 dt\,\Theta[x(t)]
\label{eq:tau}
\end{equation}
where $\Theta[x]$ is the Heaviside step function. The time $\tau$ is a random variable depending on the history of the Brownian particle. The cumulative distribution $\mathbb{P}(\tau)$ is given by the celebrated arcsine law \eref{eq:arcsin}, see e.g. \cite{Ito1965,Feller1968,Morters2010}. Equivalently, the arcsine distribution has the density
\begin{equation}
\label{arcsine}
P(\tau) = \frac{1}{\pi}\,\frac{1}{\sqrt{\tau(1-\tau)}}
\end{equation}
This formula shows that, contrary to a naive expectation, it is more likely that $\tau$ is either small or close to $1$, and it is least likely to be around $\tau=1/2$. Using \eref{eq:arcsin} or \eref{arcsine} one can compute all moments:
\begin{equation}
\langle \tau^n \rangle = 
\frac{1}{\pi}\,\frac{\Gamma\big(n+\frac{1}{2}\big)\, \Gamma\big(\frac{1}{2}\big)}{\Gamma(n+1)} = \frac{(2n-1)!!}{(2n)!!}
\end{equation}
The first moment, $\langle \tau \rangle=\frac{1}{2}$, merely reflects the right-left symmetry; the following moments, e.g.  
$\langle \tau^2 \rangle=\frac{3}{8}$ and $\langle \tau^3 \rangle=\frac{5}{16}$, cannot be deduced merely from the symmetry. 

Finding analogs of \eref{eq:arcsin} or \eref{arcsine} is of course the main challenge. Here we present a detailed computation of the second moment of $\tau$. We then briefly explain how to compute the third moment. These calculations are pedestrian and laborious, and generalizing them to the forth moment already appears very challenging.  In particular, we cannot compute $\langle \tau^n \rangle$ for large $n$ which would have shed the light on the structure of the single-file analog of the density \eref{arcsine}. 
 
Scale invariance of the Brownian motion allows one to study the behavior on an arbitrary time interval, so the common choice is $0 \le t \le 1$. In the case of single-file diffusion we must consider long intervals: $0\leq t\leq T$ with $T\gg (D\rho^2)^{-1}$. We shall measure all time variables in units of $T$, so e.g. $\tau=T_+/T$ is the fractional time that the tagged particle spends to the right of the origin. 

The average time which the tagged particle spends to the right of the origin is the same as the time it spends to the left, so $\langle \tau \rangle=1/2$. The first non-trivial  moment is $\langle \tau^2 \rangle$. This second moment is different from that of the Brownian motion. We will see that the result depends on whether we perform or not the averaging over initial conditions. The former setting is known as annealed, the latter is deterministic or quenched. 

Using the definition \eref{eq:tau} we find
\begin{equation}
\langle \tau^2 \rangle =2\int_{0}^{1}dt_2\int_{0}^{t_2}dt_1 \, 
\left\langle \Theta[X(t_1)]\Theta[X(t_2)]\right\rangle
\label{eq:tau2}
\end{equation}
To complete the computation it is sufficient to know the joint probability of positions at two times. Let $P({\bf X}, {\bf t})$ be this joint probability, where ${\bf X} = (X_1,  X_2)$ and ${\bf t} = (t_1, t_2)$. From \eref{eq:multi Gaussian} we find that $P({\bf X}, {\bf t})$ is Gaussian which can be written as
\begin{equation}
\fl \qquad P({\bf X}, {\bf t})=\frac{1}{2\pi \sigma_1\sigma_2\sqrt{1-R^2}} \exp\left[ -\frac{1}{2(1-R^2)}
\left(\frac{X_1^2}{\sigma_1^2}-2R\,\frac{X_1 X_2}{\sigma_1\sigma_2}+\frac{X_2^2}{\sigma_2^2}\right)\right]
\label{eq:bivariate}
\end{equation}
where 
\begin{equation}
\sigma_1^2=\langle X_1^2\rangle, \quad  \sigma_2^2=\langle X_2^2\rangle, \quad   R=\frac{\langle X_1 X_2 \rangle}{\sigma_1\sigma_2}
\end{equation}

These quantities are different in annealed and in deterministic settings:
 in the annealed case  we have, using \eref{SD_1d},
 $\sigma_j^2=\mathcal{D(\rho)}\,\sqrt{4t_j}$; when the initial condition is fixed
 we have, using  \eref{SD_det}, $\sigma_j^2=\mathcal{D(\rho)}\,\sqrt{2t_j}$.
 The quantity $R$ depends on the ratio $r\equiv t_1/t_2$. From  \eref{SF:2T}--\eref{SF:2T_det} we get
\begin{equation}
R(r)=\cases{\frac{1+\sqrt{r}-\sqrt{1-r}}{2\,r^{1/4}}  & annealed \\ 
\frac{\sqrt{1+r}-\sqrt{1-r}}{\sqrt{2}\,r^{1/4}}            & deterministic}
\label{eq:rho}
\end{equation}

Using these results we can compute the average $\langle \Theta[X(t_1)]\Theta[X(t_2)]\rangle$ in \eref{eq:tau2}. One can avoid calculations by relying on a neat formula \cite[p.~937]{Abramowitz1964}
\begin{equation}
\fl \qquad \langle \Theta[X(t_1)]\Theta[X(t_2)]\rangle=\int_{0}^{\infty}dX_1\int_{0}^{\infty}dX_2 P({\bf X}, {\bf t})=\frac{1}{4}+\frac{1}{2\pi}\arcsin R(r)
\label{eq:quadrant}
\end{equation}
which is valid for any bivariate Gaussian distribution \eref{eq:bivariate}. Interestingly, the final result \eref{eq:quadrant} does not depend on $\sigma_1$ and $\sigma_2$, it depends only on the Pearson correlation coefficient $R(r)$. For completeness, we note that Eq.~\eref{eq:quadrant} can be easily derived without referring to \cite{Abramowitz1964}, it suffices to compute the Gaussian integral over $X_2$ and then perform the integration over $X_1$ using an identity 
\begin{equation*}
\int_{0}^{\infty}dx\, e^{-x^2}\erfc{q\, x}=\frac{1}{\sqrt{\pi}}\arccos\left(\frac{q}{\sqrt{1+q^2}}\right).
\end{equation*}

Substituting \eref{eq:quadrant} in \eref{eq:tau2} leads to 
\begin{equation}
\label{tau:2}
\langle \tau^2 \rangle= \int_0^1 dt_2 \int_0^{t_2} dt_1\,\left[ \frac{1}{2} + \frac{1}{\pi} \arcsin R(r)\right].
\end{equation}
Recalling that $r=t_1/t_2$ we simplify \eref{tau:2} to 
\begin{equation}
\langle \tau^2 \rangle= \frac{1}{4} + \frac{1}{2\pi} \int_{0}^{1}dr\,\arcsin R(r).
\label{eq:tau2 final}
\end{equation}
This together with \eref{eq:rho} completes the calculation  of the second moment. Numerically, 
\begin{equation}
\langle \tau^2 \rangle = 
\cases
{0.353766\ldots           &      annealed\\
0.326602\ldots            &  deterministic.}
\end{equation}
Equation \eref{eq:tau2 final} applies to any stochastic process for which the joint probability distribution $P({\bf X}, {\bf t})$ is Gaussian. For instance, for standard one-dimensional Brownian motion $R(r)=\sqrt{r}$ and plugging this into \eref{eq:tau2 final} we obtain $\langle \tau^2\rangle=\frac{3}{8}$ which can be alternatively deduced from \eref{arcsine}.

It is remarkable that the final results for $\langle \tau^2 \rangle$ are universal, i.e. they do not depend on the details of the single-file system. This is a general feature, it continues to hold for the higher moments as we show below. 

Is it feasible to use the same strategy for computing higher moments? Using \eref{eq:tau} we obtain
\begin{equation}
\langle \tau^n \rangle =n!\int_{0}^{1}dt_n\cdots\int_{0}^{t_2}dt_1 \, 
\left\langle \Theta[X(t_1)]\cdots \Theta[X(t_n)]\right\rangle
\label{eq:tau3}
\end{equation}
so one should be able to compute the average $\langle \Theta[X(t_1)]\cdots \Theta[X(t_n)]\rangle$. This quantity is often referred to as the \textit{orthant probability} \cite{Bacon1963}. Some results about the orthant probabilities are known in the interesting to us situation when the joint probability distribution of $X(t_j)$'s is a multivariate Gaussian \eref{eq:multi Gaussian}. From the general structure of \eref{eq:multi Gaussian} we already conclude that the orthant probabilities depend on the correlation coefficients $R_{ij}=R(t_i/t_j)$ and do not depend on (non-universal) variances $\sigma_j^2$. Since $R(r)$ are same for all single-file systems, all the moments \eref{eq:tau3} are universal.

It is difficult, however, to establish \textit{explicit} results despite of the fact that the joint probability distribution is a multivariate Gaussian distribution. Explicit (and simple) formulas have been derived only for $n=2$, Eq.~\eref{eq:quadrant}, and for $n=3$. In this latter case \cite{Bacon1963} 
\begin{equation}
\fl \qquad \langle \Theta[X(t_1)]\Theta[X(t_2)] \Theta[X(t_3)]\rangle=\frac{1}{8}+\frac{1}{4\pi}\left[ \arcsin R_{12}+\arcsin R_{23}+\arcsin R_{13}\right]
\label{eq:orthant}
\end{equation}
Combining \eref{eq:tau3} and \eref{eq:orthant} we find
\begin{equation}
\fl \qquad \qquad \langle \tau^3 \rangle= \frac{1}{8} + \frac{3}{4\pi} \int_{0}^{1}dr\,\arcsin R(r)=\cases{0.280649\ldots           &      annealed\\
0.239903\ldots            &  deterministic.}
\end{equation}

We also note that the second moment has been recently computed \cite{Benichou2015} for the single-file system composed of Brownian point particles in a very inhomogeneous setting, namely when {\it all} particles are initially at the origin. The result is $\langle \tau^2 \rangle = \frac{3}{8}+\frac{1}{8}J_0(\pi)=0.33696\ldots$, where $J_0$ is a Bessel function. The methods of Ref.~\cite{Benichou2015} crucially rely on the composition of the system and hence they are entirely different from the methods which we use.

\section{Law of the simple logarithm}
 \label{sec:logarithm}
 
Typical displacements in Brownian motion grow as $\sqrt{t}$ with time. Of course, there are excursions beyond the typical distance. A way to quantify the likelihood of these excursions is to find the smallest envelope that contains the Brownian motion with probability one at infinite time limit. The answer is given in the law of the iterated logarithm,  Eq.~\eref{BM:log+}.   Informally, the smallest envelop grows $\sqrt{\ln(\ln t)}$ times faster than the typical displacement. 

Here we argue that for the single-file diffusion the minimum envelope grows $\sqrt{\ln(t)}$ times faster than the typical displacement. Therefore the tagged particle has relatively wider excursions than the Brownian particle.

We shall use heuristic arguments, so we first show that they lead to the well-known correct result for the Brownian motion. Let us think about the envelop as an absorbing boundary moving to the right according to
\begin{equation}
\label{BM:psi}
b(t) = \sqrt{2\sigma^2(t)\,\alpha(t)}
\end{equation}
where $\alpha(t)$ is a function that diverges as $t\to\infty$ and $\sigma^2(t)=2Dt$ is the variance. Denote by $S(t)$ the survival probability of the Brownian particle. Since the envelope moves faster than the standard deviation, we assume that the probability distribution still has a Gaussian form, but with normalization $S(t)$. We write
\begin{equation}
\label{BM:alive}
P(x,t) = \frac{S(t)}{\sqrt{2 \pi \sigma^2(t)}}\,\exp\!\left[-\frac{x^2}{2 \sigma^2(t)}\right].
\end{equation}
The survival probability has to be determined self-consistently. This comes from the condition that the rate at which the $S(t)$ decreases must be equal to the probability current at the position of absorbing envelope: 
\begin{equation}
\label{BM:St-eq}
\frac{dS}{dt} = D \frac{\partial P}{\partial x}\Big|_{x=b(t)}
\end{equation}
Plugging \eref{BM:alive} into \eref{BM:St-eq} we obtain
\begin{equation}
\label{BM:St}
\frac{dS}{dt} = -\frac{DS}{\sigma^2}\,\frac{e^{-\alpha}\,\sqrt{\alpha}}{\sqrt{\pi}}.
\end{equation}
For Brownian motion $\sigma^2(t)=2Dt$. We insert this into \eref{BM:St} and note that
 if  $e^{-\alpha}\,\sqrt{\alpha} \sim (\ln t)^{-1-\delta}$ for any $\delta>0$, then the survival probability approaches a finite value in the $t\to\infty$ limit; however,  if $e^{-\alpha}\,\sqrt{\alpha} \sim (\ln t)^{-1+\delta}$ for any $\delta\geq 0$, then the survival probability vanishes when $t\to\infty$. The borderline case is thus  obtained for
 $e^{-\alpha}\,\sqrt{\alpha} \sim (\ln t)^{-1}$, i.e. $\alpha \simeq \ln(\ln t)$. This together with \eref{BM:psi} leads to \eref{BM:log+}.

For the single-file diffusion we proceed as in the case of the Brownian motion. At large times the distribution for the tagged particle position $X(t)$ is asymptotically Gaussian, but with a variance that grows as $\sigma^2(t)\sim \sqrt{t}$. Combining this with \eref{BM:St} we find that the minimal growth rate of the envelope for which the survival probability remains asymptotically non-zero is characterized by the condition $e^{-\alpha}\,\sqrt{\alpha} \sim (\ln t)^{-1}t^{-1/2}$, from which we get
\begin{equation}
\alpha(t) \simeq \frac{1}{2}\ln t + \frac{3}{2}\ln(\ln t).
\end{equation}
The term with iterated logarithm is now sub-leading, so it is asymptotically negligible. Thus the envelop grows according to Eq.~\eref{SF:log+}. To contrast with the Brownian case, \eref{SF:log+} can be called the law of the simple logarithm. 

\section{General Formalism}
\label{sec:mft}

The trajectories of the tagged particle in single-file diffusion can be characterized at a macroscopic (hydrodynamic) scale using the macroscopic fluctuation theory  (MFT) \cite{Bertini2014}. Indeed, the single-file constraint allows one to express the displacement of the tagged particle through the density field \cite{KMS2014,KMS2015} and thereby to apply the MFT to analyze  the  dynamical properties of the tagged particle. 

Recall that the starting point of the MFT is a hydrodynamic equation (or Langevin equation) for the fluctuating macroscopic density profile $\rho(x,t)$:
\begin{equation}
	\partial_{t}\rho=\partial_{x}\left[
	D(\rho)\partial_{x}\rho+\sqrt{\sigma(\rho)}\eta\right]
	\label{eq:fhe}
\end{equation}
where $\eta(x,t)$ is a Gaussian noise with mean zero and covariance
\begin{equation}
	\langle \eta(x,t)\eta(x^{\prime},t^{\prime})\rangle=\delta(x-x^{\prime})\delta(t-t^{\prime}).
\end{equation}
The angular bracket denotes ensemble average. On the macroscopic scale, all the microscopic details of the dynamics are encapsulated into the transport coefficients $D(\rho)$ and $\sigma(\rho)$, the diffusivity and mobility, respectively \cite{Derrida2007,Bodineau2004}. 

Consider the trajectory of the tagged particle $X(t)$ in the time window $[0,T]$.  The position of the tagged particle at any time is related to the macroscopic density profile by the condition that total number of particles on either side of the tagged particle is invariant over time. This leads to 
\begin{equation}
	\int_{0}^{X(t)} \rho(x,t)dx=\int_{0}^{\infty}\left[
	\rho(x,t)-\rho(x,0) \right]dx
	\label{eq:XT definition}
\end{equation}
where we have assumed  that the tagged particle starts at the origin at $t=0$. 

Thus the tagged particle position is a functional $X(t)\equiv X_t[\rho]$ of the density profile $\rho(x,t)$.
The statistics of the trajectories $X(t)$ can be characterized in terms of the generating functional $\langle \exp[\int dt\,\lambda(t)X(t)]\rangle$. In our previous work \cite{KMS2014,KMS2015} we had $\lambda(t)=\lambda\delta(t-T)$, so we studied the generating function $\langle \exp[\lambda X(T)]\rangle$ encoding the full statistics of the displacement $X(T)$ at a single time. The general case when $\lambda(t)$ is an arbitrary function encodes the statistics over the entire time window. It is often convenient to study the logarithm of the average of the generating functional
\begin{equation}
\label{ML:functional}
\mu[\lambda,T]=\log \left \langle \exp\left[\int_0^T dt\,\lambda(t)X(t)\right] \right\rangle
\end{equation}
When $\lambda(t)=\lambda\delta(t-T)$, the function $\mu(\lambda,T)=\log \langle \exp[\lambda X(T)]\rangle$ is known as the cumulant generating function since its expansion gives all the cumulants $\langle X^n(T)\rangle_c$. Generally we shall call \eref{ML:functional} the cumulant generating functional.  
The Legendre transformation of functional $\mu[\lambda,T]$ is the action $A[X]$ characterizing the probability weight of each trajectory $X(t)$ of the tagged particle, in the large $T$ limit.

Starting with the fluctuating hydrodynamic equation \eref{eq:fhe} we write the
moment generating functional as a path integral
\begin{equation}
	\left \langle \exp\left[\int_0^T dt\,\lambda(t)X(t)\right] \right\rangle  
	= \int\mathcal{D}\left[\rho,\hat{\rho}\right] e^{-S_T[\rho,\hat{\rho}]}
\end{equation}
with action
\begin{eqnarray}
S_T[\rho,\hat{\rho}] &=& -\int_0^T dt\,\lambda(t) X_t[\rho]+F[\rho(x,0)] \nonumber\\
	&+& \int_{0}^{T}dt
	\int_{-\infty}^{\infty}dx \left\{ \hat{\rho}\partial_{t}\rho -
	\frac{\sigma(\rho)}{2}\left( \partial_{x}\hat{\rho}
	\right)^{2}+D(\rho)\left( \partial_{x}\rho \right)\left(
	\partial_{x}\hat{\rho}
	\right) \right\}.
\label{eq:action}
\end{eqnarray}
This action has been derived using a Martin-Siggia-Rose formalism \cite{Martin1973}. The derivation of \eref{eq:action} is essentially identical to the derivation \cite{KMS2014,KMS2015} of the action in the special case when $\lambda(t)=\lambda\delta(t-T)$, the only distinction is in the first term on the right-hand side of \eref{eq:action}. Hence we just  explain the meaning of $\hat{\rho}$ and $F[\rho(x,0)]$. The field $\hat{\rho}$ is a Lagrange multiplier ensuring the validity of Eq.~\eref{eq:fhe}. For the deterministic initial condition the second term on the right-hand side of \eref{eq:action} is absent, so we can set $F=0$; for the annealed setting $F[\rho(x,0)]$ is the free energy associated with the initial density profile $\rho(x,0)$. For the initial state with uniform average density $\rho$, this free energy can be written \cite{Derrida2007} as
\begin{equation}
	F\left[ \rho(x) \right]=\int_{-\infty}^{\infty}dx
	\int_{\rho}^{\rho(x)}dr \frac{2D(r)}{\sigma(r)}\left[ \rho(x)-r
	\right].
	\label{eq:F}
\end{equation}

The action $S_T[\rho,\hat{\rho}]$ grows with increasing time and at large $T$ the path integral is dominated by the path that corresponds to the least action. We denote this optimal path by $(q,p)=(\rho,\hat{\rho})$ and perform
  a small variation $\rho\rightarrow q+\delta
\rho$ and $\hat{\rho}\rightarrow p+\delta \hat{\rho}$ around this  optimal path. To the leading order, the
change in action $S_T[\rho,\hat{\rho}]$ corresponding to this variation is
\begin{eqnarray}
\fl\delta S_T= \int_{-\infty}^{\infty}dx \left\{\frac{\delta F[q(x,0)]}{\delta
	q(x,0)} -p(x,0) - \int_0^T dt \, \lambda(t)\,\frac{\delta X_{t}[q]}{\delta q(x,0)}\right\}\delta \rho(x,0)\nonumber\\
        +\int_{-\infty}^{\infty}dx~ p(x,T)~ \delta
	\rho(x,T)\nonumber\\
	+\int_{0}^{T}dt \int_{-\infty}^{\infty}dx \left\{
	-\partial_{t}p-\frac{\sigma^{\prime}(q)}{2}\left( \partial_{x}p
	\right)^{2} -D(q)\partial_{xx}p -\lambda(t)
	\frac{\delta X_{t}[q]}{\delta q(x,t)}\right\}\delta \rho(x,t)\nonumber\\
	+\int_{0}^{T}dt \int_{-\infty}^{\infty}dx \left\{
	\partial_{t}q+\partial_{x}\left(\sigma(q) \partial_{x}p
	\right) -\partial_{x}\left(D(q)\partial_{x}q\right)\right\}\delta
	\hat{\rho}(x,t)
	\label{eq:action variation}
\end{eqnarray}
where the functional derivatives are taken at the optimal path $(q,p)$.
In the above we have used the property that $X_t[\rho]$ is a functional of only the initial
profile $\rho(x,0)$ and the profile $\rho(x,t)$ at time $t$ and has
 no dependence on intermediate times.

For the action $S_T[q,p]$ to be minimum, the variation $\delta S_T$ must vanish: $\delta
S_T[q,p]=0$. Ensuring this condition we arrive at the Euler-Lagrange equations and the  boundary conditions. More precisely, since  $\delta \rho(x,t)$ and $\delta\hat{\rho}(x,t)$ are arbitrary, the vanishing of the last two integrals on the right-hand side of \eref{eq:action variation} leads to the following Euler-Lagrange equations
\begin{eqnarray}	
\partial_{t}q-\partial_{x}\left(D(q)\partial_{x}q\right)=-\partial_{x}\left(\sigma(q) \partial_{x}p
	\right)\label{eq:optimal q}\\
\partial_{t}p +D(q)\partial_{xx}p=-\frac{1}{2}\,\sigma^{\prime}(q)\left( \partial_{x}p
	\right)^{2}-\lambda(t)
	\frac{\delta X_{t}[q]}{\delta q(x,t)}.\label{eq:optimal p}
\end{eqnarray}

Using Eq.~\eref{eq:XT definition} we rewrite Eqs.~\eref{eq:optimal q}--\eref{eq:optimal p} as
\begin{eqnarray}	
\partial_{t}q-\partial_{x}\left(D(q)\partial_{x}q\right)=-\partial_{x}\left(\sigma(q) \partial_{x}p
	\right)\label{eq:optimal q2}\\
\partial_{t}p +D(q)\partial_{xx}p=-\frac{1}{2}\,\sigma^{\prime}(q)\left( \partial_{x}p
	\right)^{2}-B(t)\Theta(x-Y(t))
\label{eq:optimal p2}
\end{eqnarray}
where we have used the shorthand notation
\begin{equation}
 Y(t) =X_t[q], \qquad   B(t)=\frac{\lambda(t)}{q(Y(t),t)}.
 \label{eq:def B}
\end{equation}

The boundary conditions result from  the vanishing of the first two integrals in
\eref{eq:action variation}. In the annealed setting both $\rho(x,0)$ and $\rho(x,T)$ are fluctuating,
equivalently, $\delta \rho(x,0)$ and $\delta \rho(x,T)$ are arbitrary. Then, for
the minimal action, their coefficients in \eref{eq:action variation} must vanish, leading to the boundary condition
\begin{eqnarray}
	p(x,0)=-\int_0^Tdt~\lambda(t)\frac{\delta X_{t}[q]}{\delta q(x,0)}+\frac{\delta
	F[q(x,0)]}{\delta q(x,0)}\label{eq:p0}\\
	p(x,T)=0.
\label{eq:boundary pT}
\end{eqnarray}
Recalling \eref{eq:XT definition} and \eref{eq:F}, we rewrite \eref{eq:p0} as
\begin{equation}
\label{eq:boundary p0}
p(x,0)=\Theta(x)\int_0^T dt~B(t)+\int_{\rho}^{q(x,0)}dr\frac{2D(r)}{\sigma(r)}.
\end{equation}

For the deterministic initial condition $\delta\rho(x,0)=0$, so the first integral in \eref{eq:action variation} identically vanishes. The vanishing of the second integral in \eref{eq:action variation} yields \eref{eq:boundary pT}. 

The cumulant generating functional is given by $\mu[\lambda,T] = -S_T[q,p]$. We can further simplify \eref{eq:action} using the Euler-Lagrange equations \eref{eq:optimal q}--\eref{eq:optimal p} to yield
\begin{equation}
\fl	\qquad  \mu[\lambda,T]=\int_0^T dt\, \lambda(t) Y(t)-F[q(x,0)]-\int_{0}^{T}dt\int_{\infty}^{\infty}dx\,
	\frac{\sigma[q(x,t)]}{2}\left(\partial_{x}p(x,t) \right)^{2}.
	\label{eq:mu}
\end{equation}
In deriving \eref{eq:mu} we have taken into account the fact 
 that the derivatives $\partial_{x}p$ and $\partial_{x}q$ vanish as $x\rightarrow \pm \infty$.

As a self-consistency check we verify that for $\lambda(t)=\lambda\delta(t-T)$ we recover the results \cite{KMS2014,KMS2015} describing the tagged particle at final time. 

It is instructive to compare the Euler-Lagrange equations \eref{eq:optimal q2}--\eref{eq:optimal p2} with Euler-Lagrange equations which appear in analysis of large deviations involving single-time characteristics. Most of previous studies were devoted to single-time quantities with problems ranging from large deviations of the stationary profile \cite{Tailleur2007,Bertini2001} or integrated current \cite{Gerschenfeld2009} in diffusive systems to the melted area in an Ising quadrant \cite{KMS_interface} and the survival of the target \cite{MVK14}. In all these examples the Euler-Lagrange equations are same, viz. they are identical to \eref{eq:optimal q2}--\eref{eq:optimal p2} with $B(t)=0$. These equations are already very complicated and time-dependent solutions were found only in the simplest case of the Brownian single-file system (see e.g. \cite{KMS2014,KMS2015,Gerschenfeld2009}). In the present case when $B(t)\ne 0$ we do not know how to solve \eref{eq:optimal q2}--\eref{eq:optimal p2} even in the Brownian case. 

It is important to analyze the scaling of $\mu[\lambda,T]$ with $T$. For this, we rescale time $\tau=t/T$ and position $\xi=x/\sqrt{T}$. Substituting this in the formal solution \eref{eq:mu} and using the Euler Lagrange equations \eref{eq:optimal q2}--\eref{eq:optimal p2} along with their boundary conditions we find the scaling relation
\begin{equation}
\mu\left[\frac{\lambda(\tau)}{T},T\right]=\sqrt{T}\,\mu[\lambda(\tau),1]
\label{eq:mu scaling}
\end{equation}
which leads to the announced sub-diffusive scaling \eref{eq:kn scaling} of the cumulants.  

\section{Two-time Correlation}
\label{sec:2time}

The formalism of the previous section applies to any single-file system. It is impossible, however, to solve the Euler-Lagrange equations \eref{eq:optimal q2}--\eref{eq:optimal p2} in the general situation when the transport coefficients $D(\rho)$ and $\sigma(\rho)$ are arbitrary. The  framework presented in Sec.~\ref{sec:mft} 
is still helpful to derive some general results. In this section we derive the two-time correlation function, \eref{SD_1d}  and \eref{SD_det}, for an arbitrary single-file system.  We employ a perturbation approach \cite{Krapivsky2012} which has already proved useful in a number of problems. In the present case, we write
\begin{equation}
	\lambda(t)=\epsilon ~h(t)
\end{equation}
and treat $\epsilon$ as a small parameter. When $\epsilon=0$ the solution is $q(x,t)=\rho$ and $p(x,t)=0$. Generally we write a series expansion
\begin{eqnarray}
	q(x,t)=\rho+\epsilon\, q_{1}(x,t)+\epsilon^{2}\,
	q_{2}(x,t)+\cdots\label{eq:expansion 1}\\
	p(x,t)=\epsilon\, p_{1}(x,t)+\epsilon^{2}\, p_{2}(x,t)+\cdots.
	\label{eq:expansion}
\end{eqnarray}
The cumulant generating function can also be written as a series 
\begin{equation}
	\mu[\epsilon h,T]=\epsilon^2 \mu_2[h,T]+\cdots.
\end{equation}
The terms in odd powers of $\epsilon$ vanish because the microscopic dynamics is unbiased. From the definition of the cumulant generating functional \eref{ML:functional} we have
\begin{equation}
	\mu_n[h,T]=\frac{1}{n!}\int_0^T dt_1 ... \int_0^T dt_n~ h(t_1)...h(t_n)\,k_n(t_1,\cdots,t_n)
	\label{eq:mu2 x1x2}
\end{equation}
where $k_n(t_1,\cdots,t_n)$ is the $n$-time cumulant. The cumulants are related to the multi-time correlations. In the second order, we have $k_2(t_1,t_2)\equiv \langle X(t_1)X(t_2) \rangle$. 

Combining the series expansion \eref{eq:expansion 1}--\eref{eq:expansion} and the Euler-Lagrange equations \eref{eq:optimal q2}--\eref{eq:optimal p2}  we find
\begin{eqnarray}
	\partial_{t}q_{1}-D(\rho)\partial_{xx}q_{1}&=&-\sigma(\rho)\partial_{xx}p_{1}\,
	\label{eq:optimal q1}\\
	\partial_{t}p_{1}+D(\rho)\partial_{xx}p_{1}&=& -\frac{h(t)}{\rho}\Theta(x) \label{eq:optimal p1}
\end{eqnarray}
in the first order. In deriving Eqs.~\eref{eq:optimal q1}--\eref{eq:optimal p1} we have taken into account that
\begin{equation}
B(t)=\epsilon\, \frac{h(t)}{\rho}+\mathcal{O}(\epsilon^2). 
\end{equation}
This can be seen from the definition of $B(t)$ in \eref{eq:def B}.  

We write a similar series expansion for the function $Y(t)$ 
\begin{equation}
Y(t)=\epsilon Y_1(t)+\epsilon^2 Y_2(t)+\cdots.
\end{equation}
Using \eref{eq:XT definition}, we establish the vanishing of the zeroth order term and find
\begin{equation}
	Y_{1}(t)=\frac{1}{\rho}\int_{0}^{\infty}dx \left[q_{1}(x,t)-q_{1}(x,0)
	\right]
\label{eq:Y1}
\end{equation}

Substituting all these series expansions into \eref{eq:mu} we get
\begin{equation}
	\mu_2[h,T]= \int_0^Tdt~h(t)Y_1(t)-F_{2}-\frac{\sigma(\rho)}{2}\int_{0}^{T}dt\int_{-\infty}^{\infty}dx
	\left( \partial_{x}p_{1}\right)^{2}
	\label{eq:X2}
\end{equation}
with $F_2=0$ for the deterministic initial condition;  in the annealed case 
\begin{equation}
	F_{2}=\frac{D(\rho)}{\sigma(\rho)}\int_{-\infty}^{\infty}dx\left( q_1(x,0)
	\right)^2 
	\label{eq:F2 annealed}
\end{equation}
follows from Eq.~\eref{eq:F}. We now determine $\langle X(t_1)X(t_2) \rangle$ by solving Eqs.~\eref{eq:optimal q1}--\eref{eq:optimal p1}.

\subsection{Deterministic initial condition}

The boundary conditions are
\begin{equation}
	p_{1}(x,T)=0\qquad \textrm{and}\qquad
	q_{1}(x,0)=0. 
\label{eq:boundary linear quenched}
\end{equation}	
The solution of \eref{eq:optimal p1} subject to $p_{1}(x,T)=0$ reads
\begin{equation}
	p_{1}(x,t)=\rho^{-1}\int_t^T d\tau \int_{-\infty}^{\infty}dz~h(\tau)~\Theta(z)~g_D(z,\tau\vert x,t)
	\label{eq:dp1 quench}
\end{equation}
where
\begin{eqnarray}
\fl	\qquad\qquad 	g_D(z,\tau\vert x,t)=\frac{1}{\sqrt{4 \pi D(\rho)(\tau-t) }}\exp\left[
-\frac{(z-x)^{2}}{4D(\rho)(\tau-t)} \right]
\end{eqnarray}
for $0\le t\le \tau$. Performing the Gaussian integral over $z$ we get
\begin{equation}
	p_{1}(x,t)=\frac{1}{2\rho}\int_{t}^{T}d\tau~h(\tau) \erfc{ \frac{-x}{2\sqrt{D(\rho)\left( \tau-t\right)}} } 
\label{eq:p1 solution}
\end{equation}
 We  shall  also need $\partial_x p_1$,  given by  
\begin{equation}
	\partial_x p_1(x,t)=\frac{1}{\rho}\int_t^Td\tau~h(\tau)~g_{\tiny{D}}(0,\tau\vert x,t) 
	\label{eq:pd p}
\end{equation}

It is convenient to write $q_{1}(x,t)=-\partial_{x}\psi(x,t)$. Plugging this into \eref{eq:optimal q1} and solving the resulting equation we obtain 
\begin{equation}
	\psi(x,t)=\sigma(\rho)\int_{0}^{t}d\tau\int_{-\infty}^{\infty}dz
	~g_{D}(x,t \vert z, \tau)~\partial_zp_1(z,\tau).
	\label{eq:psi}
\end{equation}

We now  calculate  $\mu_{2}[h]_{\rm det}$ for the deterministic initial condition,  $F_{2}=0$.  Eq.~\eref{eq:X2} becomes 
\begin{eqnarray}
	\mu_{2}[h,T]_{\rm det}&=& \int_0^Tdt~\frac{h(t)}{\rho}\int_0^{\infty}dx~q_1(x,t)-\frac{\sigma(\rho)}{2}\int_{0}^{T}dt\int_{-\infty}^{\infty}dx
	\left( \partial_{x}p_{1}\right)^{2}\nonumber\\
	&=& I_1-I_2
	\label{eq:variance quench}
\end{eqnarray}
where we have  used Eq.~\eref{eq:Y1} and $q_1(x,0)=0$.

To compute $I_1$ we write again $q_1=-\partial_{x}\psi$, integrate by parts, and take into account that $\psi(x,t)$ vanishes at $x\rightarrow \infty$:
\begin{equation}
	I_1=\frac{1}{\rho}\int_0^{T}dt~h(t)~\psi(0,t).
\end{equation}
Taking $\psi(0,t)$ from Eq.~\eref{eq:psi} and $\partial_zp_1(z,\tau)$ from Eq.~\eref{eq:pd p}, we re-write $I_1$ as
\begin{equation}
\fl\qquad	I_1=\frac{\sigma(\rho)}{\rho^2}\int_0^Tdt~h(t)~\int_0^{t}d\tau\int_{\tau}^{T}dt_1~h(t_1)
	\int_{-\infty}^{\infty}dz~g_D(0,t_1\vert z,\tau)g_D(0,t\vert z,\tau).
	\label{eq:I1 form 2}
\end{equation}
Computing the Gaussian integral over $z$, we arrive at 
\begin{equation}
\frac{I_1}{\mathcal{D}(\rho)} = \int_0^Tdt~h(t)~\int_0^{t}d\tau\int_{\tau}^{T}dt_1~h(t_1)
	\frac{1}{\sqrt{t_1+t-2\tau}}
\end{equation}
where we used the shorthand notation \eref{var_X}. The integration over $\tau$ variable is performed
using identities like
\begin{equation}
 \int_{0}^{t} \frac{d\tau}{\sqrt{t_1+t-2\tau}}=\sqrt{t_1+t}-\sqrt{t_1-t} \quad {\rm when} \quad t_1>t.
\end{equation}
This leads to 
\begin{equation}
I_1= \mathcal{D}(\rho)\int_0^Tdt_1\int_{0}^{T}dt_2~h(t_1) h(t_2)
	\left[\sqrt{t_1+t_2}-\sqrt{\vert t_1-t_2\vert} \right]
	\label{eq:I1}
\end{equation}
where we denote $t_2=t$.

To compute $I_2$, we take $\partial_x p_1(x,t)$ from Eq.~\eref{eq:pd p} and obtain
\begin{equation}
\fl \qquad I_2=\frac{\sigma(\rho)}{2\rho^2}\int_0^Tdt~\int_t^{T}d\tau~\int_{t}^{T}dt_1~h(\tau)h(t_1)
	\int_{-\infty}^{\infty}dx~g_D(0,t_1\vert x,t)g_D(0,\tau\vert x,t).
\end{equation}
Changing the order of integration, $\int_{0}^{T}dt \int_{t}^{T}d\tau = \int_{0}^{T}d\tau \int_{0}^{\tau}dt$,
we find $I_2=\frac{1}{2}I_1$ once we recall Eq.~\eref{eq:I1 form 2}. Therefore
\begin{equation}
	\mu_2[h,T]_{\rm det} =\frac{\mathcal{D}(\rho)}{2}\int_0^Tdt_1\int_{0}^{T}dt_2~h(t_1) h(t_2)
	\left[\sqrt{t_1+t_2}-\sqrt{\vert t_1-t_2\vert} \right]
	\label{eq:79}
\end{equation}
Comparing this with Eq.~\eref{eq:mu2 x1x2} for $n=2$ we arrive at the announced result for the two-time correlation
\begin{equation}
\langle X(t_1) X(t_2) \rangle_{\rm det}= \mathcal{D}(\rho)\left[ \sqrt{t_1+t_2}-\sqrt{\vert t_1-t_2 \vert}\right].
\label{eq:autocor quench}
\end{equation}

\subsection{Annealed setting}

The boundary conditions for $p_1$ and $q_1$ are now given by 
\begin{equation}
\fl \qquad p_{1}(x,T)=0, \qquad\textrm{and}\qquad 
	q_{1}(x,0)=\frac{\sigma(\rho)}{2 D(\rho)}\left[ p_{1}(x,0)-\Theta(x)\int_{0}^{T}dt \frac{h(t)}{\rho} \right].
	\label{eq:boundary linear annealed}
\end{equation}
Using \eref{eq:Y1}--\eref{eq:F2 annealed} we get 
\begin{eqnarray}
\fl  \qquad \mu_2[h,T]_{\rm ann} = \frac{1}{\rho}\int_{0}^{T}dt\,h(t)\int_{0}^{\infty}dx \left[q_{1}(x,t)-q_{1}(x,0)\right] \nonumber\\
	\qquad  - \frac{D(\rho)}{\sigma(\rho)}\int_{-\infty}^{\infty}dx \left( q_1(x,0) \right)^2
	- \frac{\sigma(\rho)}{2}\int_{0}^{T}dt\int_{-\infty}^{\infty}dx \left( \partial_{x}p_{1}(x,t) \right)^{2}. 
\label{eq:variance annealed}
\end{eqnarray}

In calculations, it is  convenient to focus on $ \mu_2[h,T]_{\rm ann}-\mu_2[h,T]_{\rm det}$. For 
both  the deterministic and the annealed settings, the governing equation for $p_{1}(x,t)$ and the boundary condition on $p_{1}(x,T)$ are identical, so the solutions coincide, and are given by
 Eq.~\eref{eq:p1 solution}. The governing equation \eref{eq:optimal q1} for $q_{1}(x,t)$ is also the
 same in both cases, but the boundary conditions are different. Since Eq.~\eref{eq:optimal q1} is linear in $q_{1}(x,t)$, we write the solution as a sum
\begin{equation}
	q_{1}(x,t)=q_{i}(x,t)+q_{h}(x,t)	
\end{equation}	
where $q_{i}(x,t)$ is the solution of the inhomogeneous diffusion equation
\begin{equation}
	\partial_{t}q_{i}-D(\rho)\partial_{xx} q_{i}=-
	\sigma(\rho)\partial_{xx}p_{1} 
	\qquad\textrm{with}\qquad q_{i}(x,0)=0
\end{equation}
and $q_{h}(x,t)$ is the solution of the homogeneous equation
\begin{equation}
\label{eq:qh}
\partial_{t}q_{h}-D(\rho)\partial_{xx}	q_{h}=0
\end{equation}
subject to the boundary condition
\begin{equation}
\label{BC:qh}
q_{h}(x,0)=\frac{\sigma(\rho)}{2D(\rho)}\left[ p_{1}(x,0)-\Theta(x)\int_{0}^{T}dt \frac{h(t)}{\rho} \right].
\end{equation}

Since $q_{i}(x,t)$ and $q_{1}(x,t)$ in the deterministic case satisfy the same boundary-value problem, they are identical. Further, using \eref{eq:variance quench} and \eref{eq:variance annealed} we conclude that 
\begin{eqnarray}
\fl \qquad \quad  \mu_2[h,T]_{\rm ann}-\mu_2[h,T]_{\rm det} &=& \frac{1}{\rho}\int_{0}^{T}dt~ h(t) \int_{0}^{\infty}dx
	                                              \left[q_{h}(x,t)-q_{h}(x,0)\right]  \nonumber\\
 & - & \frac{D(\rho)}{\sigma(\rho)}\int_{-\infty}^{\infty}dx\, \left(q_{h}(x,0)\right)^2.
\end{eqnarray}
Thus the difference depends only on $q_{h}(x,t)$.
Further simplification comes from the  identity (proved in \ref{Ap:proof}):
\begin{equation}
	\frac{1}{\rho}\int_{0}^{T}dt~ h(t) \int_{0}^{\infty}dx
	\left[q_{h}(x,t)-q_{h}(x,0)\right]
	=\frac{2D(\rho)}{\sigma(\rho)}\int_{-\infty}^{\infty}dx\left(q_{h}(x,0)\right)^2.
	\label{eq:id 1}
\end{equation}
 Thus,
\begin{eqnarray}
\label{m2:long}
\fl \quad \quad \mu_2[h,T]_{\rm ann}-\mu_2[h,T]_{\rm det}&=&\frac{D(\rho)}{\sigma(\rho)}\int_{-\infty}^{\infty}dx\left(q_{h}(x,0)\right)^2\nonumber\\
&=&\frac{\sigma(\rho)}{4D(\rho)}\int_{-\infty}^{\infty}dx\left[p_{1}(x,0)-\Theta(x)\int_{0}^{T}dt~\frac{h(t)}{\rho}\right]^2
\end{eqnarray}
where in the last step we have used the boundary condition \eref{BC:qh}. We extract $p_1(x,0)$ from \eref{eq:p1 solution} and after straightforward calculations recast \eref{m2:long} into 
\begin{eqnarray}
\label{mm:long}
\fl \qquad 	\mu_2[h,T]_{\rm ann}-\mu_2[h,T]_{\rm det}
=\int_{0}^{T}dt_1\int_{0}^{T}dt_2~h(t_1)h(t_2)~\frac{\sigma(\rho)}{8\rho^{2}D(\rho)} \nonumber\\
\qquad \qquad \qquad\quad \int_{0}^{\infty}dx~\erfc{\frac{x}{\sqrt{4D(\rho)t_1}}}~\erfc{\frac{x}{\sqrt{4D(\rho)t_2}}}.
\end{eqnarray}
Finally, using the  identity \cite{Prudnikov1986}
\begin{equation}
	\int_{0}^{\infty}dy~\erfc{\frac{y}{\sqrt{t_1}}}~\erfc{\frac{y}{\sqrt{t_2}}}
	=\frac{1}{\sqrt{\pi}}\left[\sqrt{t_1}+\sqrt{t_2}-\sqrt{t_1+t_2} \, \right]
\end{equation}
we perform the integral over $x$ and reduce Eq.~\eref{mm:long} to
\begin{equation*}
\fl \quad  \mu_2[h,T]_{\rm ann}-\mu_2[h,T]_{\rm det}
=\frac{\mathcal{D}(\rho)}{2}\int_{0}^{T}dt_1\int_{0}^{T}dt_2~h(t_1)h(t_2) \left[\sqrt{t_1}+\sqrt{t_2}-\sqrt{t_1+t_2}\,\right]
\end{equation*}
which in conjunction with Eq.~\eref{eq:79}  lead to the announced result
\begin{equation}
\langle X(t_1) X(t_2) \rangle_{\rm ann}= \mathcal{D}(\rho)\left[ \sqrt{t_1}+\sqrt{t_2}-\sqrt{\vert t_1-t_2 \vert}\right].
\label{eq:autocor anneal}
\end{equation}

\section{Summary}
 \label{sec:summary}

We have investigated dynamical properties of a tagged particle in single-file diffusion. We computed the two-time correlation function for the displacement of the tagged particle. The results \eref{SF:2T}--\eref{SF:2T_det} are simple, but the derivations are involved since they rely on the macroscopic fluctuation theory (MFT). The advantage is the generality: our predictions \eref{SF:2T}--\eref{SF:2T_det} apply to an {\it arbitrary} single-file system, all the specificity of a concrete system is encapsulated in an amplitude $\mathcal{D}(\rho)$ which is a simple combination of the fundamental transport coefficients, $D(\rho)$ and $\sigma(\rho)$, characterizing the concrete single-file system. For a single-file system of Brownian particles with hardcore repulsion, our results \eref{SF:2T}--\eref{SF:2T_det} are confirmed by an independent microscopic calculation \cite{SD2015}.

Using the two-time correlation function, we calculated two non-trivial moments of the fraction of time which the tagged particle spends to the right of its starting point. The major challenge is to compute all such moments, equivalently to find the full  probability distribution. For Brownian motion, this probability distribution is governed by the arcsine law. 
We also discussed a single-file analog of the law of the iterated logarithm describing the minimal asymptotic envelope of the Brownian motion. This analog, the law of the `simple' logarithm, asserts that the ratio of maximal fluctuations of the tagged particle to typical fluctuations grows as $\sqrt{\ln t}$. 

Using the MFT we devised a formalism giving the cumulant generating functional. The cumulant generating function containing the full single-time statistics of the displacement of the tagged particle is a very special outcome of the formalism. In principle, the formalism gives the full statistics over the entire history, so all multiple-time correlation functions are encapsulated in the cumulant generating functional. As it often happens, the chief problem is that the governing Euler-Lagrange equations which one needs to solve to extract concrete results are extremely complicated. In other problems relying on the MFT, essentially the only solvable case corresponds to Brownian particles with hard exclusion constraint. In the present case even this seems to be very non-trivial, so far the solution for the generating function is known only in the single-time situation.  

 Finally, we  mention that the  Brownian motion has a number of remarkable properties, such as beautiful laws \cite{Ito1965,Feller1968,Morters2010} governing its maximum.  Single-file analogs of these properties are
 unexplored and fascinating open problems.

\bigskip\noindent
We are indebted to Davide Gabrielli for asking questions that have triggered this work. TS thanks Bernard Derrida for very useful discussions on the single-file problem. This research was partly supported by grant No.\ 2012145 from the BSF. We thank the Galileo Galilei Institute for Theoretical Physics for excellent working conditions and the INFN for partial support. 

\appendix

\section{Derivation of the identity \eref{eq:id 1}}
\label{Ap:proof}

Denote by $L$ the left-hand side of Eq.~\eref{eq:id 1}. We start by rewriting the expression as
\begin{equation}
\fl \quad	L=\int_{-\infty}^{\infty}dx\int_{0}^{T}dt\left[ \frac{h(t)}{\rho}\Theta(x) \right]q_{h}(x,t)-\int_{-\infty}^{\infty}dx\int_{0}^{T}dt\left[ \frac{h(t)}{\rho}\Theta(x) \right]q_{h}(x,0).
\end{equation}
Using the boundary condition \eref{BC:qh} and performing a rearrangement of the terms we get
\begin{equation}
L=\frac{2D(\rho)}{\sigma(\rho)}\int_{-\infty}^{\infty}dx\left[ q_{h}(x,0)\right]^2 - V
\end{equation}
with
\begin{equation}
V = \int_{-\infty}^{\infty}dx \left[ p_1(x,0)q_h(x,0) - \rho^{-1}\int_{0}^{T}dt\, h(t)\Theta(x) q_{h}(x,t)\right]
\end{equation}
To establish Eq.~\eref{eq:id 1} it suffices to show that $V=0$. Using Eq.~\eref{eq:optimal p1} we re-write $V$ as
\begin{equation*}
\fl V = \int_{-\infty}^{\infty}dx \,p_1(x,0)q_h(x,0) + 
\int_{-\infty}^{\infty}dx \int_{0}^{T}dt\, q_h(x,t)\left[\partial_tp_1(x,t) + D(\rho) \partial_{xx}p_1(x,t) \right].
\end{equation*}
Integrating $\int_{0}^{T}dt\, q_h(x,t)\partial_tp_1(x,t)$ by parts and recalling that $p_1(x,T)=0$ we get
\begin{equation}
\label{Ap:long}
	V = \int_{-\infty}^{\infty}dx\int_0^{T}dt \left[D(\rho)q_h(x,t)\partial_{xx}p_1(x,t) - p_1(x,t)\partial_t q_h(x,t) \right].
\end{equation}
Using \eref{eq:qh} we reduce \eref{Ap:long} to 
\begin{equation}
	V = D(\rho)\int_0^{T}dt \int_{-\infty}^{\infty}dx 
	~\partial_x\left[q_h(x,t)\partial_{x}p_1(x,t) - p_1(x,t)\partial_x q_h(x,t)\right]
\end{equation}
which clearly vanishes. Thus $V=0$ as we claimed. This completes the proof of \eref{eq:id 1}.

\section*{References}
\bibliographystyle{iopart-num}
\bibliography{reference}

\end{document}